# On the formation of gyration-like excitations in solid solutions

S I Morozov

Institute for Physics and Power Engineering, Bondarenko sq.1, 249033, Obninsk, Russia

E-mail: morozov@ippe.ru

**Abstract.** Measurements of thermal excitations of V-O and Y-Ba-Cu-O solid solutions have been performed by a method of inelastic neutron scattering for low-energy transfers range. Features at energy transfer about 3meV and about 5meV for V–O and Y-Ba-Cu-O accordingly were observed in neutron scattering spectra. The reason of appearance of the observed low-energy excitations are the formation of an effective potential of oxygen with broad weakly upwards bottom at the *xy* plane of tetragonal symmetry octahedron, which the interstitial atom occupies. The experimental observed features can be understood in the representation of hindered quantum gyration of the interstitial atom. It was assumed that O1 and O4 atoms in Y-Ba-Cu-O, as well as in V-O system, are in two-dimensional quantum-well, which has almost flat bottom formed by the surrounding atoms Cu and Ba. The possibility of formation of collective excitations of gyration-type as a result of exchange interaction of light atoms in matrix lattice is considered. The assumption is made that the existence of gyration-like excitations can play an important role for explanation of some phenomena in solid state physics, in particular in realization of high values of temperatures of superconducting transitions in high temperature superconductors.

**1. Introduction**

In the recent work [1] it was reported that the satellites of the vibrational local modes of oxygen in tantalum lattice were observed by method of inelastic slow neutrons scattering (INS). Side bands at energy transfer $\varepsilon \approx \omega 3 \pm 6$ meV were observed in the spectra of localized states in addition to the peaks of fundamental vibrational modes of oxygen ( $\omega_{1,2} \approx 42$ meV and $\omega_3 \approx 82$ meV). On the basis of this findings the conclusion was drawn that the satellites were connected with vibration-rotation excitations of oxygen in the field of surrounding metal atoms. The assumption was made that the peaks of related to the quantum gyration excitations of impurity atom should be expected in the INS spectra of some interstitial solid solutions. The condition of the observation of these excitations is an instability of the system relative to order-disorder or order-order phase transition.

In the present article the results of investigation and analysis of the low-energy region of thermal excitations in $VO_{0.024}$ and both tetragonal (T) and orthorhombic (O) phases of $YBa_2Cu_3O_{7y}$ solid solutions are presented. The V-O system undergo a transition from disordered α-phase to ordered α'-phase when the temperature is decreased [2]. The phase transition in $YBa_2Cu_3O_{7-y}$ from T to O has place at $y \approx 0.6$. It is well known that this phase transition is accompanied by ordering of O1 atoms in Cu1-O1 chains along *b*-axis of orthorhombic cell. The main goal of this experiments was to reveal a low energy excitations related to the quantum gyration excitations of oxygen atoms.

The INS measurement of solid solution $VO_{0.024}$ was carried out using IN4 direct-geometry spectrometer (Institute Laue-Langevin, Grenoble, France). The INS-measurements of $YBa_2Cu_3O_{7y}$ was carried out using direct-geometry spectrometer DIN-2PI [3] (Joint Institute of Nuclear Research, Dubna, Russia).

**2. Investigations of the $VO_{0.024}$ solid solution**

The INS-spectra from $VO_{0.024}$ alloy are presented in the figure 1 for different values of momentum transfer Q. Weak features were observed in the spectra range corresponding to creation of thermal excitations with energy transfer $\varepsilon \approx 1.3$ meV and $\varepsilon \approx 2.4$ meV ($\varepsilon = E_0 - E_f$, $E_0$ and $E_f$ are initial and final neutron energies respectively). The peaks position don't depend on neutron momentum transfer Q. At the same time the intensity of the peaks are increasing with the growing momentum transfer. The structure of the spectrum and the dependence of scattering intensities on momentum transfer *Q* makes it possible to assume that the observed peaks are related to the gyration-type motion of oxygen atom. The basis of this assumption follows.

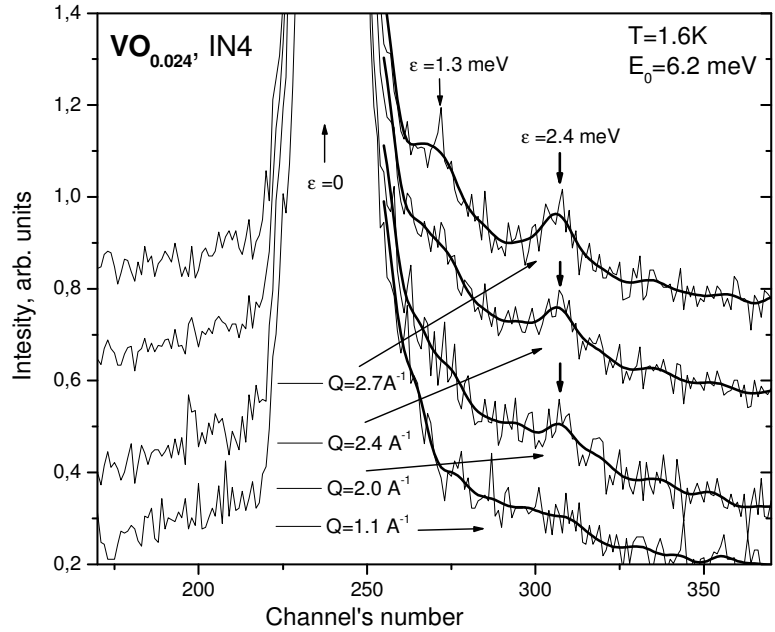

Figure 1. INS spectra of $VO_{0.024}$ solid solution at different momentum transfer of neutrons at $T=1.6K$. The peaks of low energy excitations at are shown by vertical arrows. The positions of peaks are indicated in meV.

In the V-O solid solution the oxygen occupies octahedral-sites (O-sites) (see figure 2) in vanadium. Interaction of oxygen with the nearest metal atoms carry mainly a covalent character. Interaction of oxygen with the nearest metal atoms carry mainly a covalent character. Thus the distance between metal and metalloid is determined first of all by the sum of metal radius of metal atom and covalent radius of metalloid. On the other hand, neglecting the local lattice distortions, the distance between the nearest neighboring vanadium and oxygen atoms in $VO_{0.024}$ $R_{V-O}$ is substantially smaller than the sum of the atomic radii $R_{V+O}$ [4] ($R_{V-O}= a/2=1.52$ Å and $R_{V+O}=R_V+R_O=1.34Å+0.66Å=2.0Å$). Therefore the oxygen produces a strong deformations around itself. Although the local distortions in the vicinity of an impurity atom are large (in β'-phase of V-O the displacements are about $2\Delta_1\approx0.53Å$ [5]), the V–O distance in the α-phase is likely to remain substantially smaller than the equilibrium distance $r_0\approx R_{V+O}$ of V–O pair interaction. Hence, one can expect the high negative values of the transverse bonds $f_t = V'(r_1)/r_1$, where $r_1$ is the distance between atom of oxygen and the nearest neighbors and $V'(r_1)$ is the derivative of the pair V–O interaction potential $V(r)$ at the point $r_1$ [6]. On the other hand, the distance from the center of the O-site to the second nearest neighbors $R_2 \approx a/\sqrt{2} = 2.15$ Å is noticeably larger than $R_{V+O}$. Under certain conditions (e.g. in the case of the α-α' phase transition or transition of the interstitial atom from coherent to incoherent state) this leads to instability of oxygen in the center of O-site. As a result the oxygen come to be a dynamically displaced out the center of O-site. Thus the lattice becomes unstable regarding to order-disorder phase transition. It can be supposed that the formation of localized gyration-type excitations are fundamentally related to structural instability of lattice near the α-α' phase transition [2]. Indeed, in this case the interstitial and apical atoms of Vanadium form a elongated quasi-molecule like a gyroscope. The gyroscope rotational axis passes through the center of mass of quasimolecula and is parallel to the line which connects the nearest (apical) see figure 3 Vanadium atoms.

The energy spectrum of this quasi-molecule resembles that of spectrum of quantum particle on the circle. Below the simple estimation of the energy of excitations of gyration like motion is presented.

In order to verify our assumption about existence of low energy rotation-like levels in spectra of interstitial atom the calculation of effective potential of interaction of oxygen with the nearest neighbors were carried out. Calculations were performed with Lennard-Jones pair potential ($E_{min}=300 meV$ and $r_0=2Å$) for the V-O interaction. The result is shown in figure 4. As is seen from figure 4 the potential $V_{eff}$ has a wide almost flat bottom in *xy* plane (in plane of second neighbors of oxygen atoms, see figure 2).

Thus in the *xy* plane the oxygen is located not in harmonic potential but in two-dimensional quantum-well.

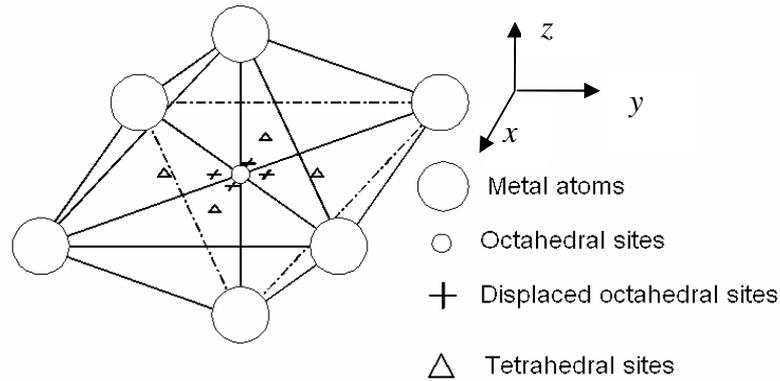

Figure 2. Octahedral interstice in a bcc lattice. The point symmetry of the position is $D_{4h}$. The shortest axis of octahedron passing through the apical atoms is coincide with z axis on this picture. Crosses indicate position of minima of effective V-O interaction potential (see text below).

In more details the effective potential $V_{eff}$ has four minima and four saddle points ( with potential barrier heights $V_4$) along the line of minima values of gradient of potential (dashed line in figure 5). The potential curve along the dashed line defines an orientational potential $V_{ang}$. An additional saddle point with greater value of energy $V_1$ is located in the center of octahedron. Thus a corrugated energy channel $V_{ang}$ around the center of O-site is formed in the *xy* plane, along which the motion of oxygen is possible.

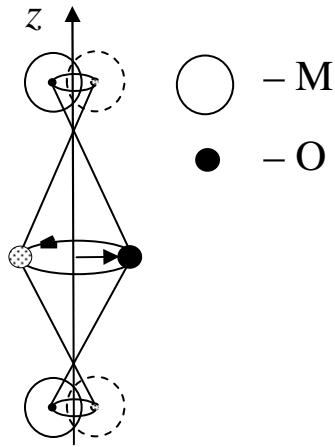

Figure 3. The interstitial atom (O) with the nearest metal (M) neighbors (two apical atoms of octahedron, see figure 2) are forming a quasi-molecule like elongated rotator.

The depth of the minima relatively the saddle points (height of potential barriers $V_1$ and $V_4$) and their displacements relatively to the center of an octahedron are depends on the displacements $\Delta_l$ of apical atoms from the position of equilibrium in perfect bcc lattice. The calculations were carried out for displacements of apical atoms $\Delta_l = 0.3$Å. These displacements lead to the formation of the quadruple energy minima at the distance $\delta r \approx 0.21$ Å from the center of O-site in $z=0$ plane (see figure **5**.). This is the displacement of interstitial atom that one should expect at the hard-sphere approximation. The potential barrier between neighboring displaced sites is $V_4 \approx 6$ meV, and between the two opposite displaced sites is $V_1 \approx 12$ meV. The height of $V_4$ is less then $\omega_b \approx 10$ meV, where $\omega_b$ is an estimation of the frequency for librations of oxygen in the potential field $V_{ang}$. That means that the motion of the oxygen in that sort of potential can be considered as a hindered gyration in *xy* plane. The potential barrier $V_4$ is not too high ($V_4 \leq \omega_b$). Then the energy levels of hindered gyration are not strongly differ from the eigenvalues of free gyration of the particle on the circle in the field of $V=const$ [7, 8].

At the approximation of free gyration the energy of the particle is $E_n = \frac{\Delta E}{2} \cdot n^2$, where $n=0,\pm 1,...$; $\Delta E = \frac{\hbar^2}{J}$; $J$ is a moment of inertia relatively to the axis of gyration. At the presence of the motion of such type the peaks of energy excitations should be observed in INS-spectrum at energy transfer of $\varepsilon = \frac{\Delta E}{2} \cdot (2n+1)$. In free-gyration approximation for radius δr=0.21Å and effective mass of the particle $m_{eff}$= 20 *atom units (au)*, the value of the constant of gyration energy transfer is $\Delta E \approx 4.5\ meV$. This result is in a satisfactory agreement with the experimental data (figure 1). In addition the dependence of the scattering intensities of observed peaks on momentum transfer $Q$ is in quality agreement with data on rotational dynamics of molecular groups (see [9,10] for example).

Then at low temperature the motion of the interstitial atom (in octahedral site with the $D_{4h}$ point symmetry) can be considered as quasi-gyration of the particle with effective mass $m_{eff}$ and angular momentum $M$.

The projection of angular momentum $M$ of particle on rotation axis can assume values $M_z = \hbar \cdot n$; $n=0, \pm 1$

Taking into account only the lowest excitation level, the angular momentum of this particle can be considered as a spin $S$ with quantum number $S$=1.

If the gyrating atom has a localized charge, then the gyration of the atom generates an additional "orbital" magnetic moment which can interpret as an intrinsic magnetic moment of the particle.

It is obvious that a lattice with interstitials atoms which have several possible close located equivalent positions, has a structural instability. The increase of impurity concentration should lead to a preferential ordering of interstitial atoms and ultimately to ordering-type phase transition. However, the situation basically is probable, that the structural instability is kept when .the "unstable" atoms forms the ordered sublattice in solid solutions.

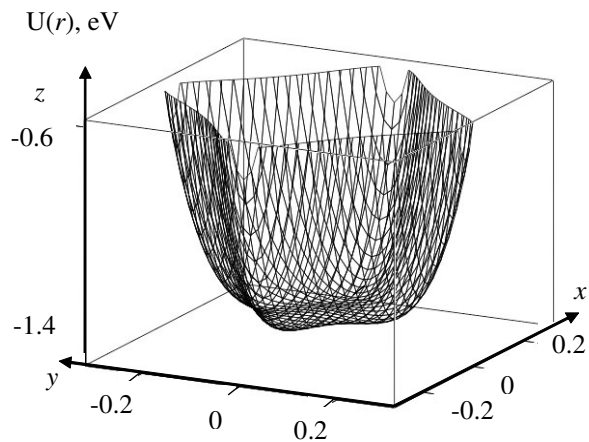

Figure 4. *XY*- plane section of effective potential of the interaction of oxygen with nearest neighbors in V-O solid solution at the region of *a-a'* phase transition. *x-y* coordinates are shown in units of $r_0$=2Å

In this case the gyration-type motion of the atoms in the corresponding sublattice may become correlated. The interaction between the gyration-type quasi-particles can result in formation of collective gyration-type excitations.

Mechanism of the formation of the collective low-energy excitation in solid solution could be presented at the quasi-classical approximation by analogy with Heisenberg model formation of the spin-wave in the magnetics. The Hamiltonian of exchange interaction for chain of the particles which have spins $S_i$ .can be written as:

$$H_{exc} = \mathcal{E}_0 - 2B \cdot \sum_{i,j=1}^{N} M_i \cdot M_j,$$

where $B$ is an exchange integral and $M_i = S_i$ is an angular momentum of the particle $i$.

The trivial solution of the equations for the motion of angular momentum is the same as for spin waves in ferromagnetics.
Corresponding law of dispersion of gyration waves for one-dimension chain of particles with antiparallel quasi-spins is given by:

$$\omega_k = 4\ |B||S||\sin(ka)|$$

where $k$ - is a wave number; $a$ – is a distance between particles in the chain.

In a case of small $k$ we have a linear dispersion law:

$$\omega \approx 4\ |B||S| \cdot a \cdot k$$

This result can be interpreted as a phased gyration of the light atoms. Thus for the antiparallel spins we obtain the same dispersion law as for phonons. Moreover, it is obvious that the intrinsic nature of this excitations is the same. On the other hand it is intuitively understandable that for excitation of gyration-like waves it is necessary to excite gyration motion of at least one atom in the chain. It means that the spectrum of gyration waves has a gap $\mathcal{E}_0$ which play the role of the energy of activation of the gyration waves. These circumstances lead to following: deformation field around an interstitial atoms (in essence it is a phonon field) provides the exchange interaction between gyration-type and phonon modes and, in this way, give rises to the formation of the hybrid gyration-type quasi-particles $\mathcal{E}_k$ with maximum of density of states in the range of energy $\mathcal{E}_0$.

In view of quantum discontinuity of the angular momentum, the energy of the excitations propagated in crystal in a form of gyration waves, will be quantized. So the gyration-type quasi-particles with energy $\mathcal{E}_k = \omega_k$ and quasi-momentum $p = k$ can be calls the gyratons. On the author's view the existing of gyration-like thermal excitations may play an important role for explanation of some phenomena in solid state physics – for instance existing a low-energy modes in superionics (e.g. see [11] for example), in metallic glasses [12,13], microscopic nature of soft modes and so on. Below we consider only one possible example of the systems where the existing of gyratons can have a fundamental importance.

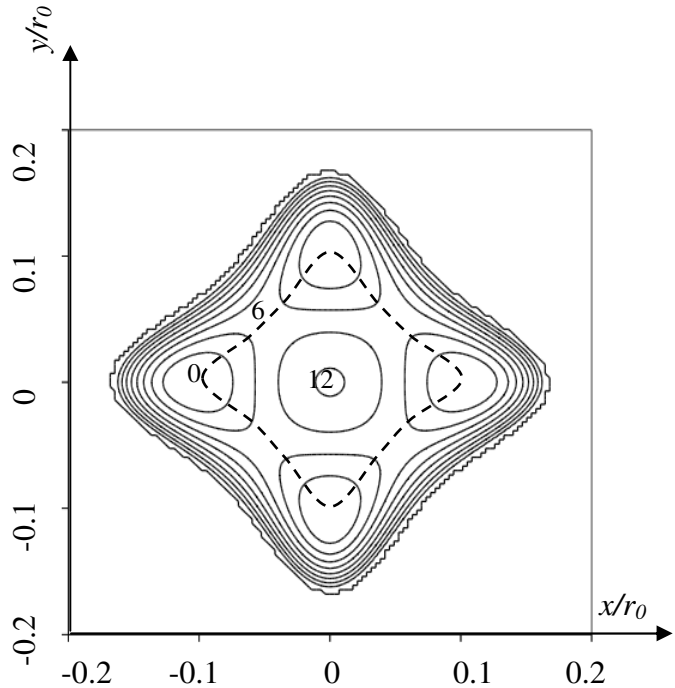

Figure 5. Map of the equipotential lines of the V–O interaction in the *x–y* plane of the octahedral site. The dashed line is the minimum gradient line of the oxygen potential energy around the center of the octahedral site (angular potential). Digits on map indicate values of potential energy of oxygen in the saddle points relative to absolute minimum in meV.

## 3. Low-frequency features in INS spectrum of Y-Ba-Cu-O

The natural candidate for systems with gyration-type collective excitations are the high temperature cuprum-oxide superconductors (HTSC) like $YBa_2Cu_3O_{7-x}$ . The reason are:

a) the structure of these compounds is defined by existing of the Cu-O chains; b) the lattices of most HTSC are related to structural instability lattices; c) The experimental data reveal large mean-square thermal displacements of oxygen O1 and O4.

In view of concept concerned in this work the large thermal displacements of the oxygen atoms are determined by their gyration-type motion. Indeed, the oxygen O1 and O4 are located in the octahedral-type surrounding of Ba and Cu atoms and the interaction of oxygen with the nearest atoms carry mainly a covalent character. Then the distance Ba-O is essentially larger, than the equilibrium one. According to the estimation the effective potential for O1 atoms has a wide weakly upwards bottom it the plane which goes through 4Ba-O1 atoms.

The situation is a little bit different for the O4 atom. However the effective potential of O4 has a relatively wide bottom in the plane which is perpendicular to *a*-axis.

The present situation is similar to the situation which has place in V-O systems near the region of order-disorder transition. Then the excited states of the oxygen at low temperature can be interpreted as a gyration-like motion of the atom around the axis passing through the Cu atoms. Correspondingly, the large experimental values of thermal displacements of the oxygen O1 and of O4 possibly is provided by their gyration-type motion.

In case of validity of the assumptions made, the low-frequency features should be expected in a density of states of YBa2Cu3O7-y at $\varepsilon \approx 3 \div 6$ meV. The same estimation of the characteristic frequency of thermal excitations follows from the data for root mean square (*rms)* thermal displacements [14-15].

In numerous investigations of vibrational density of states of Y-Ba-Cu-O, the information about so low-frequency modes are absent. An exception presents the work [16] in which the observation of low-intensity excitation at $\varepsilon \approx 5.5$ meV was reported.

In the present work a new supporting evidence was obtained concerning existence of low-energy excitations in thermal lattice spectrum of orthorhombic phase of $YBa_2Cu_3O_{6.95}$.

The results of INS-spectra measurements of Y-Ba-Cu-O in orthorhombic and tetragonal phases in room temperature are shown in figure 6.

The observed features in density of thermal states of Y-Ba-Cu-O system can't be explained in context of harmonic or anharmonic vibration approximations. On the other hand these low-energy feature at $\varepsilon \approx 5$ *meV* can testify to the existence of gyration-type motion of O1 and perhaps O4. As is seen from figure 6 the peak at $\varepsilon \approx 5$ *meV* in T-phase is noticeably wider than in O-phase. This can be explained by their different nature. In the tetragonal phase this excitation should have a localized nature. At the same time in orthorhombic phase the elementary excitations corresponding to the gyration-like motion of oxygen should have a collective nature. So in O-phase the width of the peak becomes narrowed and lower intensity because of strong dependence of momentum transfer Q.

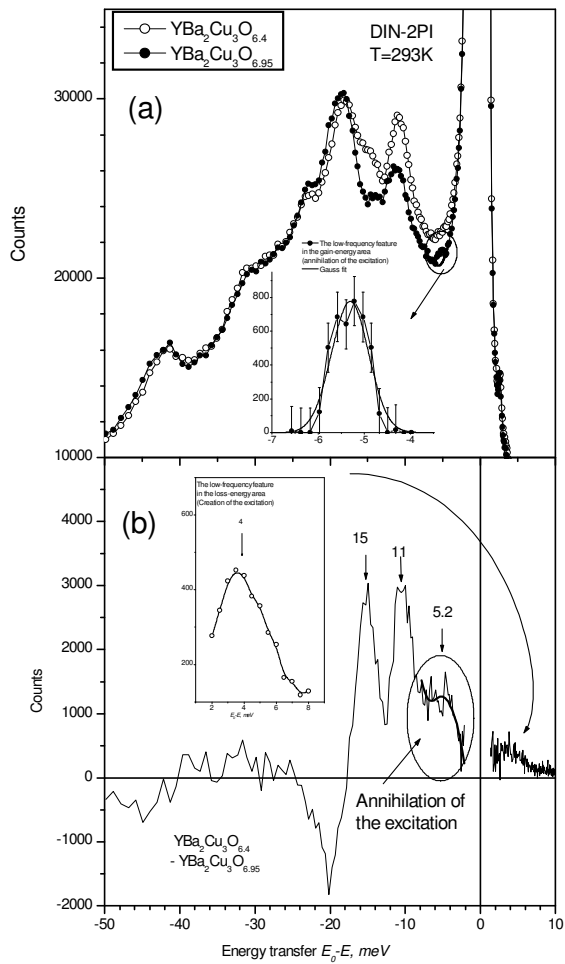

Figure 6. *a)* INS spectra of tetragonal and orthorhombic phases of $YBa_2Cu_3O_{7-x}$. Inset shows the result of the gauss-curve fitting of experimental data in the range of low-energy excitation.
*b*) Difference of spectra of tetragonal and orthorhombic phases of $YBa_2Cu_3O_{7-x}$. Inset shows a result of summarizing of experimental points in equal-distance energy scale at the low-energy range of creation of thermal excitation. Elastic neutron scattering range is excluded.

The existence of gyration-like excitations in HTSC can play principal role in the realization of mechanism of charge carriers coupling.

Because of large mean radius of quasi-gyrations one can expect a large value of a potential determining a mean energy of the charge carriers coupling and, as a result, high temperature of superconducting transition. The value of the polarization interaction between charge carriers is conditioned by the radius of oxygen gyration. On the other hand the radius is determined by the size of quantum well in which the particle is localized and almost not depends on mass of the particle.

If the charge is localized on oxygen atom, the gyration of the atom generates an additional "orbital" magnetic moment which can take part at the process of exchange interaction between spin and orbital momentum of electrons of surrounding atoms. This may lead to creation of another types of excitation (orbitons, magnons).

It seems, that the model of charge carriers coupling by exchange of virtual gyration-like excitations can explain not only high temperature of superconducting transition, but also a linear temperature contribution in heat capacity at $T<T_c$, weak isotopic dependence of $T_c$, and probably strong antiferromagnetic correlations in cuprum-oxide superconductors.

## 4. Summary

In inelastic neutron scattering spectra the peaks at energy transfer $\varepsilon \approx 1 \div 3$ meV and $\varepsilon \approx 3 \div 6$ meV for V–O and Y-Ba-Cu-O accordingly were observed for the first time. On the basis of experimental data and model consideration it is possible to derive a conclusion that presence of instability in interstitial solid solutions can result in generation of low frequency excitations of a gyration type.

In the presence of the interaction between light atoms located in the quantum well with a wide weakly upwards bottom, the gyration-type excitations can have a collective nature. The concept of existence of such excitations launched in the present work can facilitate understanding of some physical phenomena - formation of a superconducting condition and explanation of specific properties of HTSC, behavior of the density of states of superionics in low-energy range, the understanding of the microscopic nature of soft modes at order-order phase transitions in crystals and others.

## 5. Acknowledgements.

Author thank professor A.F.Gurbich for generous help in preparation of this article.